\documentclass[graybox]{svmult}

\usepackage{mathptmx}       
\usepackage{helvet}         
\usepackage{courier}        
\usepackage{type1cm}        
\usepackage{makeidx}         
\usepackage{graphicx}        
\usepackage{multicol}        
\usepackage[bottom]{footmisc}

\usepackage{amsmath}

\makeindex             

\def\[{\left[}
\def\]{\right]}
\def\({\left(}
\def\){\right)}

\newcommand{\DS}{\displaystyle}
\newcommand {\Cauchystress} {\mbox{\boldmath{$\tau$}}}  

\newcommand {\Gruneisen} {\mbox{\boldmath{$\Gamma$}}}  

\newcommand {\dfeq}{\stackrel{\mbox{\scriptsize def}}{=}}
\newcommand {\Vect} [1] {{\bf #1}}
\newcommand {\Tens} [1] {{\bf #1}}
\def\av#1{\Bigl\langle{#1}\Bigr\rangle}
\newcommand {\na}{\stackrel{\circ}{\nabla}}
\newcommand {\At}{\Vect{\tilde A}_\alpha}
\def\A{{\Vect A}}
\def\u{{\Vect u}}
\def\R{{\Vect R}}
\def\r{{\Vect r}}
\def\fa{{\Vect{F}_\alpha}}

\def\fat{{\Vect{\widetilde{F}}_\alpha}}
\def\fma{{\Vect{F}_{-\alpha}}}
\def\maal{_{-\alpha}}
\def\va{\Vect{a}_\alpha}
\def\fmat{{\Vect{\widetilde{F}}_{-\alpha}}}

\begin{document}

\title*{Equivalent thermo-mechanical parameters for perfect crystals}
\author{V.~A. Kuzkin and A.~M. Krivtsov}
\institute{V.~A. Kuzkin  \at   Institute for Problems in
Mechanical Engineering RAS, V.O., Bolshoj pr., 61 St. Petersburg, 199178, Russia,
\email{kuzkinva@gmail.com} \and
A.~M. Krivtsov \at
Institute for Problems in Mechanical Engineering RAS, V.O., Bolshoj pr.61, St. Petersburg, 199178, Russia,
\email{akrivtsov@bk.ru}}
\maketitle

\abstract{Thermo-elastic behavior of perfect single crystal is
considered. The crystal is represented as a set of interacting particles~(atoms).
The approach for determination of  equivalent continuum values for the discrete system is proposed.
Averaging of equations
of particles' motion and long wave approximation are used in order to make
link between the discrete system and equivalent continuum. Basic balance equations for equivalent continuum are derived from microscopic equations.
Macroscopic values such as Piola and Cauchy
stress tensors and heat flux are represented via microscopic
parameters. Connection
between the heat flux and temperature is discussed. Equation of state in Mie-Gruneisen form connecting Cauchy stress tensor with deformation gradient and thermal energy is obtained from microscopic considerations.}

\section{Introduction}
Determination of the relation between parameters of
discrete and continuum systems is one of the challenging problems
for modern physics.
There were intensive investigations in this area for the last several decades. However the problem is far from its final
solution.
 At the beginning the problem was only of a fundamental
interest.  However, practical interest is increasing now. The
increase is caused by fast development of discrete~\cite{kuzkin:bib:Alien} and discrete-continuum~\cite{kuzkin:bib:Wagner, kuzkin:bib:Rudd} methods
for simulation of mechanical behavior of bodies under mechanical
and thermal loadings. Various methods for transition from discrete
system to equivalent continuum are considered in literature. Long wave approximation is used
in~\cite{kuzkin:bib:Born}. The concept of quasicontinuum is proposed
in~\cite{kuzkin:bib:Kunin}. Localization functions are used in~\cite{kuzkin:bib:Hardy, kuzkin:bib:Hoover_SPAM, kuzkin:bib:Zimmerman MMS}.
These approaches give the opportunity
to spread mechanical parameters determined in lattice nodes on all
volume of the body. Decomposition of motions on slow macroscopic
and fast thermal ones is used for description of thermal properties.
There are different approaches for decomposition. In
papers~\cite{kuzkin:bib:Hardy, kuzkin:bib:Hoover_SPAM, kuzkin:bib:Zimmerman MMS, kuzkin:bib:Zimmerman MSMSE}
the decomposition of particle velocities is carried out using localization functions. As a result, dependencies of
stress tensor and heat flux on parameters of the discrete system were obtained and analyzed.
Another approach was proposed
in~\cite{kuzkin:bib:Zhou 2005}. Fourier transformation was used for
decomposition of displacements and particle velocities.
Different methods for decomposition were discussed. It was noted
that the result of the decomposition is not unique. It should
depend on characteristic time and spatial scales of the problem.

The approach based on averaging of equations of motions and application of the long wave approximation~\cite{kuzkin:bib:Born}
was proposed in papers~\cite{kuzkin:bib:Krivtsov 2007, kuzkin:bib:Krivtsov 03 CSF}. The derivation of expressions for stress tensors for ideal crystals was carried out in book~\cite{kuzkin:bib:Krivtsov 2007}. Only pair potentials were considered. Thermal motion was neglected. The influence of thermal oscillations on mechanical properties was considered in~\cite{kuzkin:bib:Krivtsov 2007, kuzkin:bib:Krivtsov 03 CSF} for one-dimensional case. The proposed approach gives an opportunity to carry out analytical derivations. In particular, the equation of state in Mie-Gruneisen form was obtained in papers~\cite{kuzkin:bib:Krivtsov 2007, kuzkin:bib:Krivtsov 03 CSF, kuzkin:bib:Kuzkin}.

In the present paper a generalization of approaches proposed
in~\cite{kuzkin:bib:Krivtsov 2007, kuzkin:bib:Krivtsov 03 CSF} for two and
three-dimensional cases is carried out. The expressions connecting Piola and
Cauchy stress tensors and heat flux with parameters of the discrete system are derived. The approach for derivation of constitutive relations for Cauchy stress tensor and heat flux is discussed.

\section{Hypotheses}

Let us consider discrete system consisted of particles, which
form the infinite ideal crystal lattice in $d$-dimensional
space~($d=1, 2$ or $3$). Crystals with simple structure are investigated only~(i.e. crystals that are invariant to
translation on any vector connecting lattice nodes). For the
sake of simplicity let the particles interact via pairwise potential
of Lennard-Jones type. Generalization of the approach discussed in the present paper for the case of multibody interatomic potentials is considered in papers~\cite{kuzkin:bib:KuzkinAPM, kuzkin:bib:KuzkinPraha}.

Two main principles are used for transition from discrete
system to equivalent continuum: the long wave approximation~\cite{kuzkin:bib:Born} and decomposition of particles' motions
 into slow continuum and fast thermal one~\cite{kuzkin:bib:Krivtsov 03 CSF, kuzkin:bib:Zhou 2005}. First let us
focus on decomposition. In literature it is carried out using
different types of averaging such as spatial averaging, time
averaging, averaging over phase space or over frequency spectrum,
etc. It was noted in paper~\cite{kuzkin:bib:Zhou 2005} that unique
decomposition is impossible because rules for a choice of
averaging parameters like averaging time, representative
volume, etc. do not exist. The only possible rule for these  parameters is that they
should depend on time and spatial scales of the problem
being solved. Let us denote average and oscillating~(thermal)
components of physical value $f$ as~$\av{f}$  and $\widetilde{f}$
respectively. Obviously,
\begin{equation}\label{kuzkin:eq:av}
 f = \av{f} + \widetilde{f}, \qquad  \widetilde{f} \dfeq  f - \av{f}.
\end{equation}
Different expressions for the averaging operator~$\av{}$ are proposed in
literature. The following operator was used in
paper~\cite{kuzkin:bib:Krivtsov 03 CSF} for one dimensional case
\begin{equation}\label{}
 \av{f_n} = \frac{1}{T \Lambda} \int_{t-T/2}^{t+T/2}
\sum_{k=n-\Lambda/2}^{n+\Lambda/2} f_k(t) dt.
 \end{equation}
where $f_k$ is magnitude of physical value $f$ for particle number $k$. Parameters $T$ and
$\Lambda$ should satisfy the following relations~$1\ll\Lambda\ll N$, $T_{min}\ll T\ll T_{max}$,
where $T_{min}$ and $T_{max}$ are minimal and maximal periods of oscillations in the system,
$N$ is the total number of particles. Obviously these limitations are too weak. Direct and inverse
Fourier transformations were used for decomposition  in paper~\cite{kuzkin:bib:Zhou 2005}. The direct transformation gives
\begin{equation}\label{}
    F(\nu) = \frac{1}{\sqrt{2\pi}} \int_{-\infty}^{\infty} f e^{i \nu t} dt,
\end{equation}
where
\begin{equation}\label{}
\av{F}(\nu) = \left\{ \begin{aligned}
F(\nu), \nu < \nu_{cutoff} \\
     0, \nu \geq \nu_{cutoff}
\end{aligned}\right.
\qquad \widetilde{F}(\nu) = \left\{ \begin{aligned}
0, \nu < \nu_{cutoff} \\
     F(\nu), \nu \geq \nu_{cutoff}
\end{aligned}\right.
\end{equation}
Here $F$ is Fourier transform of value $f$; $i$ is imaginary unit; $\nu_{cutoff}$ is cut-off frequency, which should be taken in the range 0.5-50 THz~\cite{kuzkin:bib:Born}. Inverse Fourier transformation was used in
order to obtain~$\av{f}$ and $\widetilde{f}$
\begin{equation}\label{}
 \av{f} =\frac{1}{\sqrt{2\pi}} \int_{-\infty}^{\infty}
  \av{F} e^{-i\nu t} d\nu \qquad \widetilde{f} =\frac{1}{\sqrt{2\pi}} \int_{-\infty}^{\infty}
  \widetilde{F} e^{-i\nu t} d\nu. \end{equation}
In the framework of the given approach the choice of cut-off frequency is almost arbitrary. In papers~\cite{kuzkin:bib:Hardy, kuzkin:bib:Hoover_SPAM, kuzkin:bib:Zimmerman MSMSE}
the following relations were used for decomposition of particle velocities
\begin{equation}\label{kuzkin:eq:localf}
  f_k =
  \av{f}(\Vect{x}, t) + \widetilde{f}_k(\Vect{x}, t), \qquad \av{f}(\Vect{x},t) =
  \frac{\sum_{k=1}^{M} m_k f_k \psi(\Vect{x}-\Vect{x}_k)}{\sum_{k=1}^{M}
 m_k  \psi(\Vect{x}-\Vect{x}_k)}.
\end{equation}
Here $f_k, m_k, \Vect{x}_k$ are velocity, mass and radius-vector of particle number~$k$; $\Vect{x}$ is coordinate of the spatial point where velocity is calculated; $\psi$ is localization function; $M$ is the total number of particles in the system. Decomposition~(\ref{kuzkin:eq:localf}) can be considered as spatial averaging with weight determined by function~$\psi$. Note that according to this approach thermal component of the velocity~$\widetilde{f}_k(\Vect{x}, t)$ is  continuum value (it is determined at points between particles). Several thermal velocities $\widetilde{f}_{k 1}(\Vect{x}_0, t), \widetilde{f}_{k 2}(\Vect{x}_0, t),...$ and one continuum velocity~$\av{f}(\Vect{x}_0, t)$ are simultaneously determined in one spatial point~$\Vect{x}_0$. However, formally it does not lead to any contradictions. This type of decomposition as well as all mentioned above is not unique. It strongly depends on the choice of localization function. In particular, if localization area surrounds single atom, then thermal component of the velocity is equal to zero.

Since the unique decomposition does not exist, the theory should not be based on particular method of decomposition. In addition, the results of the theory should not qualitatively change with the change of the method. Properties of particular methods for averaging are not used in the present paper, unless otherwise
stated. Let us speak about averaged~$\av{f}$ and thermal~$\widetilde{f}$ components of physical value~$f$, which are connected by formula~(\ref{kuzkin:eq:av}).

The second important statement used in the present paper is the long
wave approximation~\cite{kuzkin:bib:Born}. The idea of the approximation is the
following: an average component of any physical value is assumed to
be slowly changing in space on the distances of an order of the
interatomic distance. Then the average component can be considered as a
 continuum function of a space variable and can be expanded into
power series with respect to interatomic distance. The resulting
series should converge rapidly. Exactly this assumption allows to
make transition from a discrete system to an equivalent continuum.

\section{Kinematics}
Let us use material description of the equivalent continuum. Two configurations of continuum and discrete  system are considered: reference and actual. For the sake of simplicity let us take undeformed configuration of the crystal lattice as the reference one. Radius-vectors of equivalent continuum in the reference and actual configurations are denoted as~$\Vect{r}$ and $\Vect{R}$ respectively. Two ways of particles' identification are used. On the one hand,  the position of the particle is determined by its radius-vector. On the other hand, let us use local numbering~\cite{kuzkin:bib:Krivtsov 2007}. Starting with one reference particle let us mark all its neighbors by index~$\alpha$. Let us denote a vector connecting the reference particle with its neighbor number~$\alpha$ as~$\va$. By the definition
vectors~$\va$ have the following property
\begin{equation}\label{kuzkin:eq:vavma}
   \va = - \Vect{a}_{-\alpha}.
\end{equation}
The same vectors in an actual configuration  are represented as a sum of averaged component~$\A_\alpha$
and thermal component~$\At$. They can be expressed in terms of  vectors~$\va$ and displacements of particles as
$$
   \A_\alpha =
   \va + \u_\alpha - \u, \qquad \At =
   \widetilde{\u}_\alpha - \widetilde{\u}.
$$
Here $\u$, $\u_\alpha$ and $\widetilde{\u}$, $\widetilde{\u}_\alpha$ are average and thermal
components of displacements respectively.

The introduced way of particle identification has several properties. Let us use the following definition.
A physical value determined by the state of one particle is called a single-particle value.
For example, particle's mass, radius-vector, velocity, displacement etc.
 Let $f(\r_0)$ be one-particle value that corresponds to the reference particle with radius-vector~$\r_0$ in the reference configuration.
Denote value $f$, which corresponds to particle number $\alpha$, as~$f_\alpha(\r_0)$. Then the following two designations are equivalent
\begin{equation}\label{kuzkin:eq:t1}
  f_\alpha(\r_0) \equiv f(\r_0+\va).
 \end{equation}
In the framework of this approach the magnitude of physical
value~$f$ at the point~$\r_0$ can be represented in the following equivalent forms
\begin{equation}\label{kuzkin:eq:t2}
 f(\r_0) = f_\alpha(\r_0-\va)=f_{-\alpha}(\r_0+\va).
 \end{equation}
One can show that for multiplication of two one-particle
values~$f$ and $g$ the following identities are satisfied
\begin{equation}\label{kuzkin:eq:t3}
 \[ f_\alpha  g\](\r_0)=\[ f g_{-\alpha}\](\r_0+\va) \qquad \[ f_{-\alpha} g \](\r_0)=\[ f g_\alpha\](\r_0-\va).
\end{equation}
Hereinafter square brackets mean that all values in them are
calculated at the same point. Let us also consider values that depend on the state the reference particle and its neighbor number~$\alpha$,
notably vector connecting two particles and force acting between particles. These values  have the following property
\begin{equation}\label{kuzkin:eq:t4}
  h_\alpha(\r_0) = - h_{-\alpha}(\r_0+\va).
\end{equation}
If $h$ is a force acting between particles, then equation~(\ref{kuzkin:eq:t4}) is a specific form of Newton's third law.
The following identities are satisfied for one-particle value $f$ and the value $h$
\begin{equation}\label{kuzkin:eq:t5}
  \[ f_\alpha h_\alpha \](\r_0) = - \[f h_{-\alpha}\](\r_0+\va) \qquad~ \[f h_{-\alpha}\](\r_0) = - \[ f_\alpha h_\alpha\](\r_0-\va).
\end{equation}

Let us consider kinematics of the discrete system in the long wave approximation. It is assumed that average values of particle radius-vectors are identical to positions of corresponding points of continuum media.
Thus if some particle has radius-vector~$\r$ in the reference configuration,
then the average value of the radius-vector in actual configuration is equal to~$\R(\r)$.
The average position of its neighbor number~$\alpha$ is determined by vector~$\R(\r+\va)$.
Then one can show that vectors~$\A_\alpha$ and $\va$, connecting the particles, are related by the following formula
\begin{equation}\label{kuzkin:eq:A}
   \A_\alpha = \R(\r + \va) - \R(\r) \approx \va \cdot \na\R,
\end{equation}
where $\na$ is nabla-operator in the reference configuration. Here the long wave approximation  was used, which allows to leave the first order terms only. One can see that expression~(\ref{kuzkin:eq:A}) is similar to the formula used in continuum mechanics that connects vectors~$d\r$ and $d\R$. Using equation~(\ref{kuzkin:eq:A}) one can derive relations between vectors~$\A_\alpha$, $\va$ and  measures of deformation used in nonlinear theory of elasticity~\cite{kuzkin:bib:Lurie}. For example, the following identity fulfills for
Cauchy-Green measure~$\Tens{G}$ 
 \begin{equation}\label{}
    \A_\alpha^2 =  \va\va \cdot\cdot \Tens{G}, \qquad \Tens{G} \dfeq (\na\R)\cdot(\R\na).
 \end{equation}

\section{Equation of momentum balance}

Let us obtain the equation of motion for the equivalent continuum. Thereto  let us write down the equation of motion for the reference particle
 and use the decomposition of motions
\begin{equation}\label{kuzkin:eq:emd}
 m \ddot{\u} =
 \sum_\alpha \av{\fa(\A_\alpha + \At)},  \qquad  m \ddot{\widetilde{\u}} =
 \sum_\alpha \fat(\A_\alpha + \At),
\end{equation}
where $\fa$ is a force acting on the reference particle from its neighbor~$\alpha$; $m$ is particle's mass. The first equation from~(\ref{kuzkin:eq:emd}) describes slow motion of the system. The motion can be considered as motion of continuum media. The second equation describes thermal oscillations. One can see that both equations are coupled via the argument of the force~$\fa$. However, if the dependence of the force on the distance between particles is linear, then equations become independent. It reflects the well-known fact that harmonic models can not describe coupled thermo-mechanical effects such as thermal expansion~\cite{kuzkin:bib:Leibfrid}.
Let us conduct the following transformations in the first equation from~(\ref{kuzkin:eq:emd}).
\begin{equation}\label{kuzkin:eq:emP}
 m \ddot{\u} =
 \sum_\alpha \av{\fa} =
 \sum_\alpha \av{\fma} = \frac{1}{2}\sum_\alpha\av{\fa +\fma}.
\end{equation}
Force~$\fa$   satisfies Newton's third  law, i.e.~$\fa(\r-\va) = - \fma(\r)$~(see formula~(\ref{kuzkin:eq:t4})). Averaging
this expression and using long wave approximation one obtains
\begin{equation}\label{kuzkin:eq:focus2}
  \av{\fma}(\r) \approx -\av{\fa}(\r)+ \va\cdot\na \av{\fa}(\r)
\end{equation}
Substituting formula~(\ref{kuzkin:eq:focus2}) into equation~(\ref{kuzkin:eq:emP}) and dividing
both parts by volume of elementary cell in the reference
configuration~$V_0$ one obtains
\begin{equation}\label{kuzkin:eq:emP1}
 \frac{m}{V_0} \ddot{\u} = \na \cdot \left( \frac{1}{2V_0}\sum_\alpha \va
 \av{\fa} \right)
\end{equation}
Let us compare formula~(\ref{kuzkin:eq:emP1}) with equation of motion for
continuum in Piola's form~\cite{kuzkin:bib:Lurie}.
\begin{equation}\label{kuzkin:eq:Piolacont}
 \rho_0 \ddot{\u} = \na \cdot \Tens{P}.
\end{equation}
where $\Tens{P}$ is Piola stress tensor, $\rho_0$ is a density in the reference configuration.
Comparing equations~(\ref{kuzkin:eq:emP1}) and  (\ref{kuzkin:eq:Piolacont}) one can deduce that
\begin{equation}\label{kuzkin:eq:Piola}
  \Tens{P} = \frac{1}{2V_0}\sum_\alpha \va \av{\fa}, \qquad \rho_0 =
  \frac{m}{V_0},
\end{equation}
 Strictly speaking the first formula from~(\ref{kuzkin:eq:Piola}) is satisfied only with accuracy of tensor with zero divergency. This tensor corresponds to some equilibrium stress field in the crystal.

Let us conduct the same derivations in an actual configuration. Equation of motion for the particle has form~(\ref{kuzkin:eq:emP}).
 Let us rewrite formula~(\ref{kuzkin:eq:focus2}) in an actual configuration.
\begin{equation}\label{kuzkin:eq:focus3}
\fa(\R-\A_\alpha) = - \fma(\R) \quad\Rightarrow\quad \av{\fma}(\R) \approx
-\av{\fa}(\R)+ \A_\alpha\cdot \nabla \av{\fa}(\R).
\end{equation}
Here it was used that in the long wave approximation~$\A_{-\alpha}\approx -\A_{\alpha}$. Substituting  expression~(\ref{kuzkin:eq:focus3}) into equation~(\ref{kuzkin:eq:emP}) and
dividing both parts by the volume of elementary cell in the actual
configuration~$V$ one obtains
\begin{equation}\label{kuzkin:eq:emC}
 \frac{m}{V} \ddot{\u} =\frac{1}{2V}  \sum_\alpha \A_\alpha \cdot \nabla
 \av{\fa}.
\end{equation}
Let us conduct the following transformations in the right side of formula~(\ref{kuzkin:eq:emC})
\begin{equation}\label{}
\frac{1}{2V}  \sum_\alpha \A_\alpha \cdot \nabla\av{\fa} =
\nabla \cdot \left( \frac{1}{2V}  \sum_\alpha \A_\alpha \av{\fa}\right) -
\sum_\alpha \nabla \cdot \left( \frac{1}{2V}\A_\alpha \right) \av{\fa}
\end{equation}
The second term in the right side of the given equation  can be
written down in the following form using equation~(\ref{kuzkin:eq:A})
\begin{equation}\label{kuzkin:eq:Piolaid}
 \sum_\alpha \nabla \cdot \left( \frac{1}{2V}\A_\alpha \right) \av{\fa} =
 \frac{V_0}{2} \sum_\alpha \nabla \cdot \left( \frac{V_0}{V} \left( \R\na\right)   \right) \cdot \va \av{\fa} = 0,
 \end{equation}
where Piola's identity~$\nabla\cdot\left( \frac{V_0}{V}\left( \R\na\right)\right)\equiv~0$ was used~(see, for example,~\cite{kuzkin:bib:Lurie}).
Then equation of motion~(\ref{kuzkin:eq:emC}) has the following form
\begin{equation}\label{kuzkin:eq:emÑ2}
 \frac{m}{V} \ddot{\u} = \nabla \cdot \left( \frac{1}{2V}  \sum_\alpha \A_\alpha\av{\fa} \right).
\end{equation}
The requirement of equivalence of discrete and continuum systems
leads to the following expressions for Cauchy stress tensor and
density in the actual configuration
\begin{equation}\label{kuzkin:eq:Cauchy}
 \Cauchystress =
 \frac{1}{2V}\sum_\alpha \A_\alpha \av{\fa}, \qquad
 \rho = \frac{m}{V}.
\end{equation}
If thermal motion is not taken into account, then expression~(\ref{kuzkin:eq:Cauchy}) coincides with expressions derived in papers \cite{kuzkin:bib:Krivtsov 2007, kuzkin:bib:Zhou 2003}.

It is known that Cauchy stress tensor is symmetrical in systems without moment interactions. Let us consider tensor~$\Cauchystress$ determined
by formula~(\ref{kuzkin:eq:Cauchy}). Force~$\fa$ can be represented as
\begin{equation}\label{kuzkin:eq:Phi_def}
\fa = - \Phi_\alpha((\A_\alpha + \At)^2)(\A_\alpha + \At),   \qquad   \Phi(A^2) \dfeq -\frac{\Pi'(A)}{A}
\end{equation}
Substituting the given expression into formula~(\ref{kuzkin:eq:Cauchy}) one obtains
\begin{equation}\label{kuzkin:eq:Cauchy 2}
  \Cauchystress = -\frac{1}{2V} \sum_\alpha \av{\Phi_\alpha}\A_\alpha\A_\alpha - \frac{1}{2V} \sum_\alpha \A_\alpha
  \av{\widetilde{\Phi}_\alpha\At}.
\end{equation}
The first tensor in the right side of formula~(\ref{kuzkin:eq:Cauchy 2}) is symmetrical indeed. However the symmetry of the second tensor is not evident. Further it will be shown that antisymmetrical part of this tensor is small with respect to symmetrical part.

\section{Equation of angular momentum balance}

It is known from continuum mechanics~\cite{kuzkin:bib:Palmov 1976} that the symmetry of Cauchy stress tensor follows from equation of  angular
 momentum balance  for elementary volume. In the  discrete case elementary cell plays the role of elementary volume.
  Let us write down the averaged equation of angular momentum balance  for elementary cell~(moments are
  calculated  with respect to the center of the cell determined by vector~$\R$).
\begin{equation}\label{kuzkin:eq:sm}
     m\av{ \widetilde{\u} \times \dot{\widetilde{\u}}}\dot{\vphantom{\widetilde{\u}}} = \av{\widetilde{\u} \times \sum_\alpha \fat}
     =
     -\sum_\alpha \av{\At \times  \fat} + \sum_\alpha \av{ \widetilde{\u}_{\alpha} \times\fat}.
\end{equation}
Transforming the second term in the right side of the given equation using the long wave approximation one obtains
\begin{equation}\label{}
\begin{array}{l}
\av{\widetilde{\u}_{\alpha} \times \fat }(\R)  = - \av{\widetilde{\u} \times \fmat }(\R + \A_{\alpha}) \approx -\av{\widetilde{\u} \times \fmat} - \A_{\alpha} \cdot \nabla \av{\widetilde{\u} \times \fmat}.
  \end{array}
\end{equation}
%
%
%
Let us substitute the result into equation~(\ref{kuzkin:eq:sm}) and resolve it with
respect to~$\av{\At \times \fat}$.
\begin{equation}\label{kuzkin:eq:45}
 \frac{1}{2}\sum_\alpha\av{\At \times \fat} =
 \frac{1}{2}\sum_\alpha \A_\alpha\times \av{\widetilde{\Phi}_\alpha \At}=
 -\frac{1}{2}\sum_\alpha \A_{\alpha} \cdot \nabla \av{\widetilde{\u} \times \fmat} -
 m\av{ \widetilde{\u} \times \dot{\widetilde{\u}}}\dot{\vphantom{\widetilde{\u}}}.
  \end{equation}
Using expression~(\ref{kuzkin:eq:Cauchy 2}) for the stress tensor one can transform
formula~(\ref{kuzkin:eq:45}) to the following form
\begin{equation}\label{kuzkin:eq:sta}
 \Tens{E} \cdot\times \Cauchystress^{A}  =
  \frac{1}{2V}\sum_\alpha \A_{\alpha} \cdot \nabla \av{\widetilde{\u}\times\fmat} + \rho \av{\widetilde{\u}\times \ddot{\widetilde{\u}}}.
 \end{equation}
Here $A$ denotes  antisymmetrical part of the tensor. One can see that if there is no thermal motion, then~$\Cauchystress^A \equiv 0$. Let us show that in general case $\Cauchystress^A$ is small  in comparison with~$\Cauchystress^S$. The first term in formula~(\ref{kuzkin:eq:sta}) is small in the long wave approximation. Let us
show that if operator~$\av{}$ contains spatial averaging, then the second term is also small.
Consider the following identity
\begin{equation}\label{kuzkin:eq:kin}
 \rho \av{\widetilde{\u} \times \dot{\widetilde{\u}}}\dot{} =
 \rho \av{(\R + \widetilde{\u}) \times \dot{\widetilde{\u}}}\dot{\vphantom{\widetilde{\u}}}.
 \end{equation}
The right side of formula~(\ref{kuzkin:eq:kin}) is a derivative of angular momentum, which corresponds to thermal motion. Angular
momentum is calculated with respect to the origin of coordinates. Let
the averaging operator include spatial averaging over
significantly large volume and let us assume that thermal motion
does not lead to macroscopic rotation of the volume. Then
expressions~(\ref{kuzkin:eq:kin}) are equal to zero. As a result $\Cauchystress^A$ has
the same order as terms that were neglected in long wave
assumption. Consequently, tensor~(\ref{kuzkin:eq:Cauchy}) can be considered as approximately symmetrical.

Thus averaging operator proposed above can not be arbitrary. It
should include spatial averaging. Otherwise one can not prove the symmetry of tensor~$\Cauchystress$ determined by formula~(\ref{kuzkin:eq:Cauchy}).

\section{Equation of energy balance}

For the sake of simplicity let us assume that volumetrical forces and volumetrical heat sources  are equal to zero.
Derivations are carried out in the reference configuration. In this case averaged specific total energy per volume~$V_0$  has the following form
\begin{equation}\label{kuzkin:eq:fe}
 \rho_0{\cal E} =  \frac{1}{2}\rho_0\av{(\dot{\u} + \dot{\widetilde{\u}})^2} +
 \frac{1}{2V_0}\sum_\alpha\av{\Pi(\A_\alpha + \At)},
\end{equation}
where ${\cal E}$ is particle's total energy divided by the
mass, i.e. discrete analog for mass density of the energy. Let us
 introduce the following designations
\begin{equation}\label{kuzkin:eq:total E}
\begin{array}{l}
 \DS \rho_0{\cal E} = \rho_0\left({\cal K} + {\cal U}\right), \\[4mm]
 \DS \rho_0{\cal K} = \frac{1}{2}\rho_0\dot{\u}^2, \qquad \rho_0{\cal U} = \frac{1}{2}\rho_0 \av{\dot{\widetilde{\u}}^2} + \frac{1}{2V_0}\sum_\alpha{\av{\Pi(\A_\alpha +
 \At)}}.
 \end{array}
\end{equation}
Values ${\cal K}$ and ${\cal U}$  correspond to mass densities of macroscopic kinetic and internal energies. Calculating derivatives of kinetic and potential energies taking into account
formulas~(\ref{kuzkin:eq:emd}), (\ref{kuzkin:eq:Piola}) one can obtain  
\begin{equation}\label{kuzkin:eq:Kkk}
\begin{array}{l}
\displaystyle \frac{1}{2}\rho_0 \frac{d}{dt}\left( \dot{\u}^2 +
\av{\dot{\widetilde{\u}}}^2\right) = \rho_0\left( \dot{\u}\cdot\ddot{\u} +
\av{\dot{\widetilde{\u}}\cdot\ddot{\widetilde{\u}}}\right) = \left(\na\cdot\Tens{P}\right)\cdot\dot{\u} +
\frac{1}{V_0}\sum_\alpha \av{\dot{\widetilde{\u}}\cdot \fat}, \\[4mm]
\displaystyle \frac{1}{2V_0} \frac{d}{dt} \sum_\alpha{\av{\Pi(\A_\alpha +
  \At)}}= \frac{1}{2V_0}\sum_\alpha{\av{\fa\cdot(\dot{\A}_\alpha +
  \dot{\widetilde{\A}}_\alpha)}}
  \end{array}
 \end{equation}
where formula~$\displaystyle \fa =\frac{d\Pi}{d\A_\alpha}$ was used.
Let us conduct the following transformations
\begin{equation}\label{kuzkin:eq:Uuu2}
 \displaystyle \sum_\alpha{\av{\fa}\cdot\dot{\A}_\alpha} =
 \sum_\alpha{\av{\fa}\cdot(\dot{\u}(\r+\va) -
 \dot{\u}(r))} \approx \sum_\alpha{\va \av{\fa} \cdot\cdot \dot{\u}\na}
 \end{equation}
Summarizing expressions~(\ref{kuzkin:eq:Kkk}) and taking into account formulas~(\ref{kuzkin:eq:Piola}), (\ref{kuzkin:eq:Uuu2}) one obtains the
expression for derivative of the total energy with respect to time
\begin{equation}\label{kuzkin:eq:bE}
  \rho_0 \dot{{\cal E}} = \na\cdot\left(\Tens{P}\cdot\dot{\u} \right) + \frac{1}{2V_0}\sum_\alpha
\av{ \fat\cdot \left( \dot{\widetilde{\u}}_\alpha + \dot{\widetilde{\u}} \right) }.
  \end{equation}
Using equation of momentum balance one can show that
$\DS \rho_0\dot{{\cal K}} = \left( \na \cdot \Tens{P} \right) \cdot \dot{\u}$.
Substituting this expression into equation~(\ref{kuzkin:eq:bE}) one obtains
\begin{equation}\label{kuzkin:eq:bE1}
   \rho_0\dot{{\cal U}} = \Tens{P} \cdot\cdot \left(\dot{\u} \na\right) +
  \frac{1}{2V_0}\sum_\alpha\av{\fat\cdot(\dot{\widetilde{\u}}_\alpha + \dot{\widetilde{\u}})}.
 \end{equation}
Comparing the last expression with energy balance equation for a continuum media~\cite{kuzkin:bib:Palmov 1976}
one can conclude that expression for divergency of heat flux in the reference configuration~$\Vect{h}$ has form
\begin{equation}\label{kuzkin:eq:0}
  \na \cdot \Vect{h} = -\frac{1}{2V_0}\sum_\alpha\av{\fat\cdot(\dot{\widetilde{\u}}_\alpha + \dot{\widetilde{\u}})} = -\frac{1}{2V_0}\sum_\alpha\av{\fat\cdot\dot{\widetilde{\u}}_\alpha} -\frac{1}{2V_0}
  \sum_\alpha\av{\widetilde{\Vect{F}}\maal\cdot\dot{\widetilde{\u}}}.
 \end{equation}
Let us represent the right side of this expression in the form of
divergency. Using the first identity form~(\ref{kuzkin:eq:t5}) in the right side of formula~(\ref{kuzkin:eq:0}) one obtains
\begin{equation}\label{kuzkin:eq:2}
\begin{array}{l}
   \av{\fat\cdot\dot{\widetilde{\u}}_\alpha }(\r) =
   -\av{\fmat\cdot\dot{\widetilde{\u}}}(\r+\va),~~
    \av{\fmat\cdot\dot{\widetilde{\u}}}(\r) =
   -\av{\fat\cdot\dot{\widetilde{\u}}_\alpha}(\r-\va).
 \end{array}
 \end{equation}
Substituting formulas~(\ref{kuzkin:eq:2}) into
formula~(\ref{kuzkin:eq:0}) and applying the long wave approximation one can obtain
\begin{equation}\label{}
 \na \cdot \Vect{h} = \na \cdot \left( -\frac{1}{2V_0}\sum_\alpha \va \av{\fat\cdot\dot{\widetilde{\u}}_\alpha} \right)
 = \na \cdot \left( -\frac{1}{2V_0}\sum_\alpha \va \av{\fat\cdot\dot{\widetilde{\u}}} \right).
\end{equation}
Using this expression one can write down three representations for
heat flux in the reference configuration\footnote{Note that heat flux is determined with the accuracy of vector with zero divergency.}
\begin{equation}\label{kuzkin:eq:hhh}
\begin{array}{l}
\DS  \Vect{h} =
-\frac{1}{4V_0}\sum_\alpha \va\av{\fat\cdot(\dot{\widetilde{\u}}_\alpha+\dot{\widetilde{\u}} )} =
-\frac{1}{2V_0}\sum_\alpha\va \av{\fat\cdot\dot{\widetilde{\u}}_\alpha} =
-\frac{1}{2V_0}\sum_\alpha\va \av{\fat\cdot\dot{\widetilde{\u}}}
  \end{array}
\end{equation}
Analogous formulas can be obtained for heat flux in the
actual configuration~$\Vect{H}$
\begin{equation}\label{kuzkin:eq:HH}
\begin{array}{l}
\DS  \Vect{H} = -\frac{1}{4V}\sum_\alpha \A_\alpha
\av{\fat\cdot(\dot{\widetilde{\u}}_\alpha+\dot{\widetilde{\u}} )} = -\frac{1}{2V}\sum_\alpha
\A_\alpha \av{\fat\cdot\dot{\widetilde{\u}}_\alpha} = -\frac{1}{2V}\sum_\alpha \A_\alpha
\av{\fat\cdot\dot{\widetilde{\u}}}
  \end{array}
\end{equation}
Here formula~$\Vect{H} = \frac{V_0}{V}\left(\R \na \right) \cdot \Vect{h}$ relating heat fluxes in different
configurations was used~\cite{kuzkin:bib:KondaurovFortov}.
Note that different expressions for~$\Vect{h}$ and $\Vect{H}$ in formulas~(\ref{kuzkin:eq:hhh}), (\ref{kuzkin:eq:HH}) are equal with accuracy of terms, which were neglected in the long wave approximation.

\section{Constitutive relations for stress tensor and heat flux}
Expressions~(\ref{kuzkin:eq:Cauchy}), (\ref{kuzkin:eq:total E}) connecting micro- and macro-parameters allow to derive nonlinear constitutive
relations~(equations of state) for thermo-elastic behavior of the crystal. This problem
is considered in detail in works~\cite{kuzkin:bib:Krivtsov 2007, kuzkin:bib:Kuzkin}. Only main ideas and results are shown below. The relation between
 ``cold'' component of Cauchy stress\footnote{Stress in the crystal in the absence of thermal motion.}~$\Cauchystress_0$ and Cauchy-Green measure of deformation can be obtained
substituting formulas~(\ref{kuzkin:eq:A}),  (\ref{kuzkin:eq:Phi_def}) into formula~(\ref{kuzkin:eq:Cauchy})~(see \cite{kuzkin:bib:Krivtsov 2007} for details).
\begin{equation}\label{kuzkin:eq:eos Krivtsov}
 \Cauchystress_0 = -\frac{1}{2V_0 \sqrt{|\Tens{G}|}} (\R\na) \cdot \( \sum_{\alpha} \Phi(\va\va \cdot\cdot
 \Tens{G})\va\va \) \cdot (\na \R),
\end{equation}
where $|\Tens{G}|$ is a determinant of tensor~$\Tens{G}$. Equations of state connecting thermal component of Cauchy stress~$\Cauchystress_T \dfeq \Cauchystress - \Cauchystress_0$ with thermal energy were obtained in paper~\cite{kuzkin:bib:Kuzkin}.  The expansion of
Cauchy stress and internal energy with respect to~$\At$ was
conducted. In particular, in the first approximation the following
system was obtained
\begin{equation}\label{kuzkin:eq:sist}
\begin{array}{l}
    \DS \Cauchystress_T =-\frac{1}{2V}\sum_{\alpha} \[
        2\varPhi'_{\alpha}\A_{\alpha} \Tens{E} \A +
        \varPhi'_{\alpha}\A_{\alpha}\A_{\alpha}\Tens{E}+
        2\varPhi''_{\alpha}\A_{\alpha}\A_{\alpha}\A_{\alpha}\A_{\alpha} \] \cdot\cdot \,\av{ \At \At },
 \\ [6mm]
    \DS U_T  =-\frac{1}{2} \sum_{\alpha} \[    \varPhi_{\alpha}\Tens{E}+   2\varPhi'_{\alpha}\A_{\alpha}\A_{\alpha} \] \cdot\cdot
    \,\av{ \At \At },~~\varPhi^{(n)}_\alpha \dfeq \frac{d^{n} \varPhi}{d (A_\alpha^2)^{n}} 
 \end{array}
 \end{equation}
Here $U_T$ is a thermal energy per unit volume. It was assumed that
 $\av{\At\At} =  \frac1d\,\kappa^2\Tens{E},~~\kappa^2 \dfeq \av{\At^2}$ in order to close system~(\ref{kuzkin:eq:sist}).
 In the framework of the assumption system~(\ref{kuzkin:eq:sist}) takes form
\begin{equation}\label{kuzkin:eq:eos Kuzkin}
  \DS \Cauchystress_T =\frac{1}{V} \Gruneisen U_T, \qquad \Gruneisen \dfeq  \frac{\sum_{\alpha}\( (d+2)\varPhi'_{\alpha} + 2\varPhi''_{\alpha} A_\alpha^2 \)
    \A_{\alpha}\A_{\alpha}}{\sum_{\alpha} \( d\,\varPhi_{\alpha} + 2\varPhi'_{\alpha} A_\alpha^2 \)}.
 \end{equation}
Expression~(\ref{kuzkin:eq:eos Kuzkin}) is generalized Mie-Gruneisen
equation, where $\Gruneisen$ is tensor Gruneisen coefficient. Note that more accurate equations of state can be obtained leaving higher order terms in expansions~(\ref{kuzkin:eq:sist}). For further details about the approach for derivation of equations of state see paper~\cite{kuzkin:bib:Kuzkin}.


Let us consider propagation of small thermal disturbances. Assume that the amplitude of thermal oscillations is small in
comparison with interatomic distance. Expanding expression for heat flux~(\ref{kuzkin:eq:HH}) with respect to~$\At$ and leaving
terms of order of~$\At^2$ only one obtains
\begin{equation}\label{kuzkin:eq:HH1}
 \Vect{H} =  \frac{1}{2V}\sum_{\alpha} \( \Phi(\A_{\alpha}^2)\A_{\alpha}\Tens{E} + 2\Phi'(\A_{\alpha}^2)\A_{\alpha}\A_{\alpha}\A_{\alpha}\)\cdot\cdot
 \av{\At\dot{\widetilde{\u}}_{\alpha}},~~\Phi'\dfeq \frac{d \Phi}{d A_{\alpha}^2}.
\end{equation}
Expression~(\ref{kuzkin:eq:HH1}) is satisfied for arbitrary nonlinear elastic deformations. Let us consider the case when discrete system is free from
internal mechanical loads and constraints. In this case
deformations are caused  by thermal expansion only. Then the reference and
actual configurations approximately coincide. Linearizing expression~(\ref{kuzkin:eq:HH1}) assuming that~$\A_{\alpha} \approx \va$ one obtains
\begin{equation}\label{kuzkin:eq:hf1}
 \Vect{H} = \sum_{\alpha} {}^3\Tens{C}_{\alpha} \cdot\cdot \av{\At\dot{\widetilde{\u}}_{\alpha}} = \sum_{\alpha} {}^3\Tens{C}_{\alpha} \cdot\cdot
 \av{\At\dot{\widetilde{\u}}},
\end{equation}
where ${}^3\Tens{C}_{\alpha} \dfeq \frac{1}{2V_0}\(
\Phi(a_{\alpha}^2)\va\Tens{E} + 2\Phi'(a_{\alpha}^2)\va\va\va \)$.
Let us represent the expression for heat flux in the form of
divergency.
\begin{equation}\label{kuzkin:eq:h_f}
 \begin{array}{l}
\DS  \Vect{H} =  \sum_{\alpha} {}^3\Tens{C}_{\alpha} \cdot\cdot \av{\At\dot{\widetilde{\u}}_{\alpha}}
 \DS \approx \nabla \cdot \(\frac{1}{2} \sum_{\alpha} \va {}^3\Tens{C}_{\alpha} \cdot\cdot
\av{\widetilde{\u}\widetilde{\u}}\dot{} \) - \sum_{\alpha} {}^3\Tens{C}_{\alpha}\cdot\cdot
\av{\widetilde{\u}\dot{\widetilde{\u}}_{\alpha}}^{S}.
 \end{array}
 \end{equation}
Here the following identity was
used~$\sum_{\alpha}{}^3\Tens{C}_{\alpha} = 0$. From formula~(\ref{kuzkin:eq:h_f}) it follows that, in contrast to classical Fourier law, the heat flux depends on the set of
symmetrical tensors~$\av{\widetilde{\u}\widetilde{\u}}, \av{\widetilde{\u}\dot{\widetilde{\u}}_{\alpha}}^{S}$. Let us try to connect heat flux with temperature.
Classical ideal gas definition of temperature is used
\begin{equation}\label{kuzkin:eq:kT}
  dkT = m\av{\widetilde{\dot{\u}}^2},
\end{equation}
where $k$ is Boltsman constant. Equation~(\ref{kuzkin:eq:kT}) can be
transformed taking into account equation of motion~(\ref{kuzkin:eq:emd})
\begin{equation}{}
 dkT = m\av{\widetilde{\u}\cdot
  \dot{\widetilde{\u}}}\dot{} - \av{\sum_{\alpha}
  \fat\cdot\widetilde{\u}}
  \approx
  m\av{\widetilde{\u}\cdot
 \dot{ \widetilde{\u}}}\dot{} + \sum_{\alpha} \( \Phi\Tens{E}+ 2\Phi'\va\va \) \cdot\cdot \av{\widetilde{\u} \At}.
\end{equation}
Here the expansion into series with respect to~$\At$ is carried out.  The second order terms are leaved only. Using the definition of tensor~${}^3\Tens{C}_{\alpha}$ let us write
down the resulting system for connection between heat flux and
temperature.
\begin{equation}\label{kuzkin:eq:EOSforhf}
\begin{array}{l}
 \DS \Vect{H} = \nabla \cdot \(\frac{1}{2} \sum_{\alpha} \va {}^3\Tens{C}_{\alpha} \cdot\cdot
\av{\widetilde{\u}\widetilde{\u}}\dot{} \) - \sum_{\alpha} {}^3\Tens{C}_{\alpha}\cdot\cdot
\av{\widetilde{\u}\dot{\widetilde{\u}}_{\alpha}}^{S}\\[4mm]
\displaystyle  dkT = \frac{m}{2}\Tens{E}\cdot\cdot\av{\widetilde{\u}\widetilde{\u}}\ddot{}
 +
  \sum_{\alpha} \frac{2V_0}{a_{\alpha}^2} \va\cdot {}^3\Tens{C}_{\alpha}  \cdot\cdot \av{\widetilde{\u}\widetilde{\u}_{\alpha}-\widetilde{\u}\widetilde{\u}}^S.
  \end{array}
 \end{equation}
According to system~(\ref{kuzkin:eq:EOSforhf}) the thermal state at the given
point is determined by symmetrical tensors\footnote{Tensors $\av{\widetilde{\u}\dot{\widetilde{\u}}_{\alpha}}^S$ can be represented via time derivatives of tensors~ $\av{\widetilde{\u}
\widetilde{\u}_{\alpha}}^S$ using long wave approximation.}  $\av{\widetilde{\u}\widetilde{\u}}, \av{\widetilde{\u}
\widetilde{\u}_{\alpha}}^S$.
In general, these tensors are independent. Therefore system~(\ref{kuzkin:eq:EOSforhf}) is not closed.
\section{Concluding remarks}
The generalization of the approach proposed in~\cite{kuzkin:bib:Krivtsov 2007} that allows to carry out the transformation from discrete system to
equivalent continuum was presented. Two main principles were used
for the transformation: the decomposition of particles' motions into
continuum and thermal parts, and the long wave
assumption~\cite{kuzkin:bib:Born}. The review of different methods for
decomposition was given. It was shown that all of them contain
uncertain parameters. Therefore, the result of decomposition is
principally nonunique. Thus, one can conclude that derivations
should not be based on any specific decomposition type. The connection between kinematics of discrete system and kinematics equivalent continuum was
analyzed. Equivalent
Cauchy-Green measure of deformation for discrete system was introduced.  The transition form single particle's equation of motion to equation of motion
for equivalent continuum was carried out. Expressions connecting
Cauchy and Piola stress tensors with parameters of the discrete system
were derived. It was shown that discrete analog of Cauchy stress
tensor can be non-symmetrical. Spatial averaging is necessary for
the symmetry of this tensor. Thus, averaging operator cannot be
arbitrary and should contains spatial averaging. The energy balance equation for discrete system was considered. The equation
was transformed to the form similar to energy balance equation for a continuum system. As a result, the expression
connecting heat flux with parameters of discrete system was
obtained.  Propagation of small thermal disturbances in
undeformed crystal was analyzed. It was shown that thermal state at the point is
determined by the set of independent symmetrical tensors~$\av{\widetilde{\u}\widetilde{\u}}, \av{\widetilde{\u}
\widetilde{\u}_{\alpha}}^S$. This fact
does not allow to connect heat flux with temperature. Equation of state in generalized Mie-Gruneisen form connecting Cauchy stress tensor with deformation gradient and thermal energy is obtained from microscopic considerations.

\begin{acknowledgement}
    This work was supported by Russian Foundation for Basic Research (grants No. 08-01-00865a, 09-05-12071-ofi-m).
\end{acknowledgement}

\end{document}